\documentclass[prl,twocolumn,showpacs]{revtex4}
\usepackage{bm}
\usepackage{amstext}
\usepackage{rotating}
\usepackage{graphicx}
\usepackage{epsfig}
\usepackage{amssymb}

\begin{document}

\title{Bragg spectroscopy of a strongly interacting Fermi gas}

\author{G. Veeravalli, E. Kuhnle, P. Dyke, and C. J. Vale}

\affiliation{ARC Centre of Excellence for Quantum-Atom Optics and Centre for Atom Optics and Ultrafast Spectroscopy,
Swinburne University of Technology, Melbourne, 3122, Australia}

\date{\today}

\begin{abstract}
We present a comprehensive study of the Bose-Einstein condensate to Bardeen-Cooper-Schrieffer (BEC-BCS) crossover in fermionic $^6$Li using Bragg spectroscopy.  A smooth transition from molecular to atomic spectra is observed with a clear signature of pairing at and above unitarity.  These spectra probe the dynamic and static structure factors of the gas and provide a direct link to two-body correlations.  We have characterised these correlations and measured their density dependence across the broad Feshbach resonance at 834 G.
\end{abstract}


\maketitle
%

Strongly interacting ultracold Fermi gases provide new opportunities for the study of pairing and superfluidity.  Magnetic field Feshbach resonances allow precise tuning of the interactions between fermions in different spin states and this has led to the realisation of long lived molecular Bose-Einstein Condensates (BECs) \cite{jochim03,greiner03,zwierlein03,bourdel04,partridge05} and fermionic condensates \cite{regal04,zwierlein04}.  Since their first realisation, the properties of these gases in the BEC-BCS (Bardeen-Cooper-Schrieffer) crossover have attracted much attention \cite{ohara02,kinast05,bartenstein04,kinast04,chin04,zwierlein05,zwierlein06,partridge06}.  Recent experiments using radio frequency (rf) spectroscopy have been been particularly fruitful yielding the pair size \cite{schunck08}, low energy excitation spectrum and pairing gap \cite{stewart08,schirotzek08}.

One tool that has not yet been applied to these systems is Bragg spectroscopy.  This differs from rf spectroscopy in two important ways.  Firstly, Bragg scattering does not change the internal atomic states so there are no final state interaction shifts.  Secondly, Bragg scattering can transfer momentum larger than the Fermi momentum, $\hbar k_F$, whereas the momentum transferred by an rf photon is negligible $(\ll \hbar k_F)$.  As such Bragg spectroscopy can probe a wide region of the excitation spectrum.

Several theoretical studies of Bragg scattering in Fermi gases have predicted signatures related to pairing and superfluidity \cite{buchler04,bruun06,combescot06,challis07}.  Bragg spectroscopy has previously been used to characterise the dynamic and static structure factors of bosonic condensates \cite{stamper99,steinhauer02}.  These quantities are linked to the two-body correlation function $g^{(2)}(r)$ via the Fourier transform \cite{pines,ozeri05}.  In an interacting two component Fermi gas, strong pair correlations may exist and these can be probed with Bragg spectroscopy.  Recent Bragg experiments have studied boson/boson molecules near a Feshbach resonance \cite{aboshaeer05}, Ramsey interferometry with noninteracting fermions \cite{marzok08} and molecular condensation \cite{inada07}, but no work has yet investigated correlations in the BEC-BCS crossover.

In this letter, we report Bragg scattering experiments around the broad $s$-wave Feshbach resonance at 834 G in an ultracold gas of $^6$Li.  Bragg spectra reflect the composition of the gas, being dominated by bosonic molecules below the Feshbach resonance, pairs and free fermionic atoms near unitarity, and free fermions above resonance.  Our spectra show features present in the dynamic structure factors calculated by Combescot \textit{et al.} \cite{combescot06}.  Strong pair correlations are observed at unitarity which decay as the density is lowered highlighting the many body nature of pairing on the BCS side of a Feshbach resonance.

Our experiments use a gas of $^6$Li prepared in an equal mixture of the $|F = 1/2, m_F = +1/2\rangle$ and $|1/2, -1/2\rangle$ states evaporatively cooled in a single beam optical dipole trap \cite{fuchs07}.  This yields a highly degenerate cloud with $N = 1.3 \times 10^5$ atoms in each state at a magnetic field of 835 G.  The field is then ramped in 100 ms to a desired value and held there for both Bragg scattering and imaging.  The imaging light is tuned to be resonant with atoms in the $|1/2, 1/2\rangle$ state at each magnetic field.  On the BEC side of the Feshbach resonance, molecules can be imaged using the same frequency light as atoms as their binding energy is much lower than the transition linewidth.

All experiments presented here were performed at the lowest temperature achievable on our experiment.  At unitarity we quantify this by the empirical temperature $\tilde{T}$ obtained from one-dimensional fits of an ideal Fermi gas profile to our measured density distributions integrated over two dimensions \cite{kinast05}.  $\tilde{T}$ provides an estimate of the true temperature via $\tilde{T} \simeq T/(T_F \sqrt{1+\beta})$, where $T_F = E_F/k_B$ is the Fermi temperature and $\beta = -0.58$ is a universal constant \cite{giorgini07,kinast05}.  We find $\tilde{T} = 0.1$ and $T/T_F \sim 0.07$.  Temperatures of clouds at other magnetic fields may be calculated from the entropy \cite{hu06,haussmann07}.  Far on the BEC side of the resonance the temperature can be estimated from bimodal fits to expanded molecular density profiles.  At 700 G our measured condensate fraction of $\sim 90\%$ is consistent with $T/T_F < 0.1$ at unitarity, following an isentropic magnetic field sweep \cite{hu06}.

Bragg scattering is achieved using two counter-propagating laser beams with a small tunable frequency difference $\delta$ to create a moving periodic potential.  A two photon Bragg event transfers momentum $\hbar q = 2 \hbar k_L$ to the scattered particle.  Ignoring Doppler and interaction shifts, the Bragg resonance frequency is given by $\delta = 2 \hbar k_L^2 / m$, where $k_L$ is the wavevector of the laser and $m$ is the mass of the scattered particle.  In a Fermi gas near a Feshbach resonance, these particles can be atoms (with mass $M$), tightly bound molecules or correlated pairs (mass $2M$).  Bragg scattering therefore allows one to distinguish molecules/pairs from atoms.  The relative fractions of each depends on the temperature, $T/T_F$, and interaction parameter, $1/k_F a$, where $k_F = (48N)^{1/6} \sqrt{M \bar{\omega} / \hbar}$, $\bar{\omega}$ is the geometric mean trapping frequency and $a$ is the $s$-wave scattering length.  For $^6$Li the Bragg resonant frequencies are $\delta_{2M}/2\pi = 147$ kHz for molecules and $\delta_M/2\pi = 294$ kHz for atoms.

\begin{figure}
\centerline{\epsfig{file=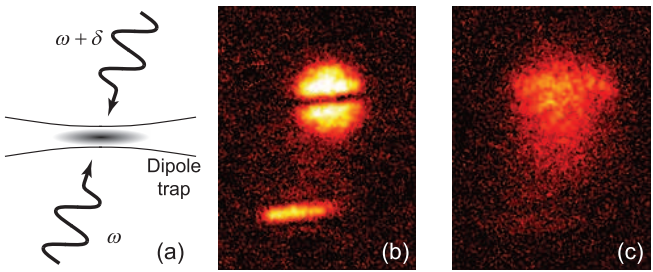,width=86mm}}
\caption{(a) Schematic of Bragg scattering experiments.  Bragg scattering of molecules from (b) a pre-expanded molecular BEC and (c) a trapped molecular BEC, both at 750 G.  The field of view for images (b) and (c) is 650 $\mu$m by 485 $\mu$m.  Elastic collisions between scattered and unscattered particles distort the resultant cloud in the trapped case.}
\label{fig:molnatoms}
\end{figure}

The Bragg lasers are applied near perpendicular to the long axis of the dipole trap, Fig. \ref{fig:molnatoms} (a), and are linearly polarised parallel to the magnetic field direction.  To demonstrate free particle scattering, Fig. \ref{fig:molnatoms} (b) shows an absorption image of molecules scattered from a molecular BEC ($\delta/2\pi = 140$ kHz) at 750 G which has expanded for 3 ms before the Bragg pulse.  The pulse duration $\tau_{Br}$, and intensity were chosen to yield a $\pi$ pulse for resonant molecules.  A thin slice of the cloud has been removed demonstrating the high momentum selectivity of Bragg scattering.  With longer Bragg pulses we have observed several Rabi cycles of scattered molecules (and atoms).

Bragg scattering of pre-expanded low density clouds however, cannot reveal the effects of interactions, so we now focus on scattering of trapped gases.  This can be complicated by elastic collisions between scattered and unscattered particles \cite{papp08}.  These result in distorted and asymmetric atomic distributions in which it is generally not possible to discern a spatially separated scattered cloud.  For Bragg spectroscopy of trapped gases, we apply a short Bragg pulse ($\tau_{Br}$ = 40 $\mu$s) and switch off the trapping potential immediately following the pulse.  Both of these events happen on a timescale short compared to the inverse trapping frequencies ($f_r$ = 320 Hz, $f_z$ = 24 Hz).  The cloud then expands for 4 ms before being imaged.  Fig. \ref{fig:molnatoms} (c) shows the outcome of this sequence for a trapped molecular BEC at 750 G.

To analyse such images, we quantify the effect of the Bragg pulse using the centre of mass displacement $X$.  The  momentum imparted to the cloud $P(q,\delta)$ by a Bragg pulse with wavevector $q$ and energy $\hbar \delta$ determines the resultant centre of mass velocity $\frac{dX(q,\delta)}{dt} = P(q,\delta)/m$.  For $q > k_F$ and times short compared to the inverse trapping frequencies this is given by \cite{brunello01,blakie02,ozeri05},
\begin{equation}\label{eqn:spec}
    \frac{d X(q,\delta)}{dt} = \frac{\hbar q \Omega_{Br}^2}{2 m} \int S(q,\delta') \frac{1-\mathrm{cos}[(\delta-\delta')\tau_{Br}]}{(\delta-\delta')^2} d \delta',
\end{equation}
where $\Omega_{Br} = (\Gamma^2/4 \Delta) \sqrt{I_1 I_2} / I_{s}$ is the two-photon Rabi frequency, $I_1$ and $I_2$ are the intensities of the two Bragg beams, $I_s$ is the saturation intensity, $\Gamma$ is the natural linewidth of the transition and $\Delta$ is the Bragg laser detuning.  $\Delta/2 \pi \sim 1.5$ GHz is chosen to be large compared to the $80$ MHz splitting between the $|1/2,1/2\rangle$ and $|1/2,-1/2\rangle$ states so the coupling to each state differs by only $5\%$.  $S(q,\delta)$ is the dynamic structure factor which describes the response of the gas to excitations.  $\Omega_{Br}^2$ is twice as large for molecules/pairs as it is for atoms (molecules have twice the polarisability) however, this is exactly compensated for in (\ref{eqn:spec}) by the factor of two mass difference.  Thus $X(q,\delta)$ has the same factor before the integral for both atomic and molecular (paired) gases.  Additionally, elastic collisions preserve centre of mass motion, so $X(q,\delta)$ is a good measure of the Bragg signal throughout the BEC-BCS crossover, independent of cloud shape and scattering length.

\begin{figure}
\centerline{\epsfig{file=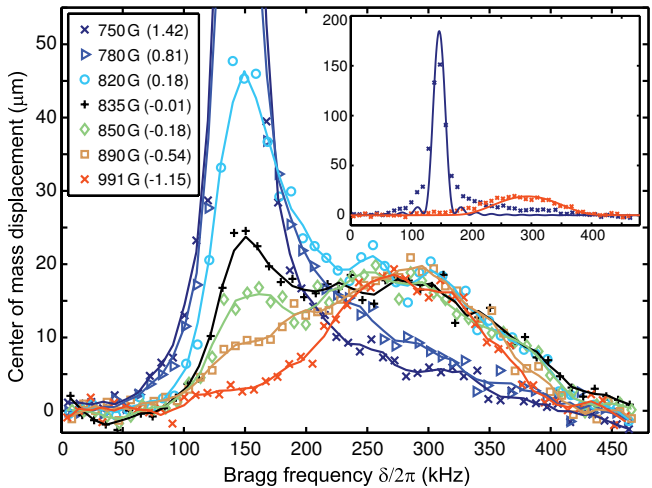,width=86mm}}
\caption{Bragg spectra showing $X(q,\delta)$ for trapped Fermi gases across the BEC-BCS crossover.  Magnetic field and ($1/k_F a$) for each spectrum are given in the legend.  The inset shows the 750 G and 991 G spectra along with the calculated $X(q,\delta)$ for an ideal Fermi gas and molecular BEC at 750 G.}
\label{fig:spectra}
\end{figure}

We evaluate $X(q,\delta)$ by integrating images such as Fig. \ref{fig:molnatoms} (c) horizontally and finding the centre of mass of the resultant line profile.  This is taken relative to the centre of mass of a reference image obtained with no Bragg pulse.  Fig. \ref{fig:spectra} shows Bragg spectra obtained in this way for a range of magnetic fields on either side of the Feshbach resonance.  At 750 G and 780 G ($1/k_F a$ = 1.4 and 0.8, respectively) the spectra are dominated by a large peak near 150 kHz corresponding to scattering of molecules from a molecular condensate.  In this regime the molecule size is smaller than the inverse excitation momentum, $a < 1/q$, so molecules behave like single particles of mass $2M$.  The height of this peak is indicative of a large condensate fraction.  The spectra are slightly asymmetric showing some excitation of free atoms.  Closer to resonance at 820 G ($1/k_F a = 0.2$) the molecule/pair peak decreases as $a \gtrsim 1/q$ and the spectra merge with the continuum of free atomic excitations.  At unitarity (835 G) the spectrum has significant components due to both pair and atom scattering.  This provides clear evidence of a substantial paired fraction in a trapped unitary gas.  At higher magnetic fields (850 G, $1/k_F a = -0.2$) the pair signal remains strong.  For 890 G and 991 G, $1/k_F a$ = -0.5 and -1.2 respectively, the pairing feature drops off and the spectra approach that of an ideal Fermi gas.  These spectra show the smooth transition in $S(q,\delta)$ from being dominated by molecular excitations below the Feshbach resonance to atomic excitations above resonance, in good agreement with the predictions of ref. \cite{combescot06}.

The inset of Fig. \ref{fig:spectra} shows the full 750 G and 991 G spectra along with the calculated $X(q,\delta)$ for a $T = 0$ ideal Fermi gas and Thomas Fermi molecular BEC ($a_{mol}$ = 110 nm at 750 G).  These limiting cases are found using the impulse approximation for $S(q,\delta)$, valid for large $q$ \cite{zambelli00}.  The bosonic molecular condensate response is much narrower and higher peaked than the Fermi gas response.  The relative heights of the calculated responses are set by requiring the area under the bosonic curve to be twice that of the fermionic curve \cite{combescot06}.  Both the heights and widths of the experimental data show good agreement with the limiting case theory when $|1/k_F a| > 1$.  The width of the molecular peak is limited by the Fourier width of the Bragg pulse ($\sim 20$ kHz).

The integral of $X(q,\delta)$ from equation (\ref{eqn:spec}) over $\delta$ is proportional to the true static structure factor, $S(q)= \hbar N^{-1} \int S(q,\delta) d\delta$ \cite{steinhauer02,ozeri05}.  Experimentally, if $\tau_{Br}$ is fixed and $\Omega_{Br}^2$ is known for a series of experiments, $S(q)$ may be obtained to within a scaling factor by directly integrating a Bragg spectrum.

\begin{figure}
\centerline{\epsfig{file=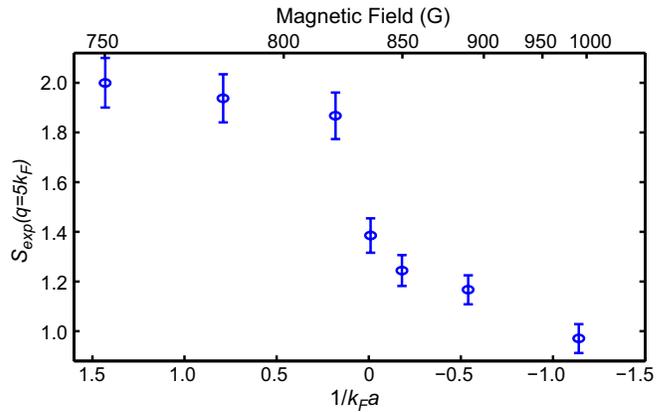,width=86mm}}
\caption{Experimental static structure factor, $S_{exp}(q=5k_F)$, as a function of $1/k_F a$ and magnetic field (top).}
\label{fig:static}
\end{figure}

In two component Fermi gases $S(q)$ is determined by correlations between particles in all combinations of spin states.  Labelling the two states with $\uparrow$ and $\downarrow$ arrows, $S(q) = S_{\uparrow \uparrow}(q) + S_{\uparrow \downarrow}(q)$, where $S_{\uparrow \uparrow}(q) = S_{\downarrow \downarrow}(q)$ for $N_\uparrow = N_\downarrow$ and $S_{\uparrow \downarrow}(q) = S_{\downarrow \uparrow}(q)$ \cite{combescot06}.  Low energy excitations ($q < k_F$) are suppressed by Pauli blocking and $S_{\uparrow \uparrow}(q \rightarrow 0) \rightarrow 0$.  However, in our experiments $q/k_F \simeq 5$ ($>$ 1) and particles in the same state are essentially uncorrelated so $S_{\uparrow \uparrow}(q) \rightarrow 1$.  The total static structure factor is then $S(q) = 1 + S_{\uparrow \downarrow}(q)$ which provides a direct link to correlations between spin-up/spin-down particles via the Fourier transform.  For a uniform gas,
\begin{equation}\label{eqn:statcorr}
    S_{\uparrow \downarrow}(q) = n \int [g_{\uparrow \downarrow}^{(2)}(r)-1] \mathrm{e}^{i q r} dr,
\end{equation}
where $n$ is the density and $g_{\uparrow \downarrow}^{(2)}(r)$ represents the spin-up/spin-down two-body correlation function \cite{combescot06}.  Quantum Monte Carlo simulations for a homogeneous gas predict $S(q > k_F)$ to vary monotonically from 2 to 1 as $1/k_F a$ goes from being large and positive to large and negative \cite{combescot06}.  In Fig \ref{fig:static} we show the experimentally determined static structure factors, $S_{exp}(q=5k_F)$, obtained by integrating the spectra in Fig. \ref{fig:spectra} over $\delta$.  These are normalised so that $S_{exp}(q=5k_F)$ at 750 G ($1/k_F a = 1.4$) is two, corresponding to the bound molecule limit.  $S_{\uparrow \downarrow}(q)$ decays from the BEC to BCS sides of the Feshbach resonance due to the decay of $g_{\uparrow \downarrow}^{(2)}(r)$.

For positive $1/k_F a$ the measured $S_{exp}(q=5k_F)$ approaches the ideal molecular value of 2 closer to the Feshbach resonance than predicted for a uniform gas \cite{combescot06}.  Our experiments employ a nonuniform trapped gas and $k_F$ refers to the local Fermi wavevector at peak cloud density.  Away from the trap centre $k_F(r) = [6 \pi^2 n(r)]^{1/3}$ decreases leading to a higher local interaction parameter $1/k_F(r) a$.  For a uniform gas, however, $k_F$ and $1/k_F a$ are constant across the cloud.  Applying equation \ref{eqn:statcorr} to a trapped gas using a local density approximation would involve averaging over regions with higher $1/k_F(r) a$ so a trapped gas would approach ideal molecular limit more rapidly than a uniform gas of equal $1/k_F a$.

\begin{figure}
\centerline{\epsfig{file=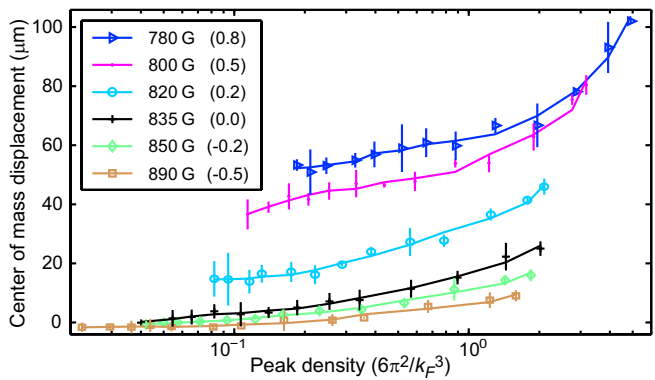,width=86mm}}
\caption{Response of trapped gases to a Bragg pulse resonant with pairs ($\delta/2\pi$ = 145 kHz) as a function of cloud density.  Magnetic field values and ($1/k_F a$) are listed in the legend.}
\label{fig:pairdens}
\end{figure}

The existence of pairs at unitarity and above a Feshbach resonance is a quantum many body effect due to the strong interactions between the two species.  As the density is lowered, interactions decrease and pair correlations decay.  To investigate this we apply a Bragg pulse resonant with pairs/molecules ($\delta/2\pi = 145$ kHz) to clouds which have been released from the trap and allowed to expand for a variable time. Imaging is always performed 4 ms after the Bragg pulse and a reference image of an unscattered cloud is obtained for each pre-expansion time.  The measured centre of mass displacements are plotted in Fig. \ref{fig:pairdens}.  Peak densities $n_0$ were obtained from the reference images and $n_0 \times (6 \pi^2 / k_F^3)^{-1}$ can be larger than the ideal gas value of one because of interactions.

On the BEC side of the Feshbach resonance the scattered pair signal starts high and decreases towards an asymptotic value as the density is lowered.  During expansion mean field energy is converted into kinetic energy and fewer molecules are resonant with the Bragg pulse \cite{stenger99}.  A nonzero asymptotic value, when collisions are negligible, implies the scattered particles are real bound molecules.  As the magnetic field is increased towards the Feshbach resonance, this asymptotic level decreases due to the stronger repulsive interactions.  At unitarity, the scattered signal starts high and tapers off eventually dying away to zero at densities below $\sim 0.1 (6 \pi^2 / k_F^3$).  This indicates that no bound molecules are present, however, in higher density clouds, pair correlations clearly exist.  Similar behaviour is observed at 850 G and 890 G but the initial high density pair signal is lower.  This technique allows us to distinguish between true bound molecules and paired atoms through the existence/nonexistence of a nonzero low density scattered signal.

In summary, we have presented the first Bragg spectroscopic study of a Fermi gas in the BEC-BCS crossover.  Bragg spectroscopy allows a direct probe of two-body correlations and the transition from molecular to atomic behaviour has been observed.  At unitarity and above the Feshbach resonance, pair scattering is observed.  Pairing is seen to decay as the density is lowered, highlighting the many body nature of pair correlations.  In future work, Bragg scattering could measure the temperature dependence of pairing and condensation \cite{inada07}.  The large momentum transfer used here prevents us from resolving the pairing gap.  Howver, low (tunable) $q$ Bragg scattering could probe the superfluid excitation spectrum and the form of $g_{\uparrow \downarrow}^{(2)}(r)$ throughout the crossover.

We thank M. Vanner for experimental assistance and  M. J. Davis, H. Hu and X.-J. Liu for helpful comments.  This project is supported by the Australian Research Council Centre of Excellence for Quantum-Atom Optics.

\bibliographystyle{amsplain}

\end{document}